\documentclass[11pt,twoside]{article}
\usepackage{asp2010}
\usepackage{natbib}

\resetcounters

\aspvolume{471} 
\aspvoltitle{Origins of the Expanding Universe: 1912-1932}
\aspcpryear{2013} 
\aspvolauthor{Michael J. Way and Deidre Hunter, eds.} 

\begin{document}

\title{V.~M. Slipher and the Development of the Nebular Spectrograph}
\author{Laird~A.~Thompson
\affil{Astronomy Department, University of Illinois, Urbana, IL, 61801, USA}}

\begin{abstract}
Vesto Melvin Slipher was the first astronomer to clearly define the factors that
determine the ``speed" of a nebular spectrograph.  This brief historical summary
recounts the way these ideas developed and how Slipher's early work on galaxy
Doppler shifts was so quickly extended in the 1930s when Milton Humason and
Edwin Hubble at Mt. Wilson Observatory began to push the velocity-distance
relationship to such a depth that no one could doubt its cosmological
significance.  \end{abstract}

\section{Early Spectroscopes and Spectrographs}

Starting with the invention of the optical spectroscope in 1814 by Joseph
Fraunhofer, and continuing for many years into the era of the first astronomical
spectrographs, no one gave serious consideration to the consequences of
preferentially selecting the \textit{f}/ratio--or equivalently the focal
length--of the last imaging lens in the system (the camera lens in a modern
spectrograph).  We know today that a spectrograph with a short focal length
camera is key to detecting extended low surface-brightness objects, i.e.
nebulae, and that longer focal length cameras work well on higher
surface-brightness objects like stars, the Sun and planets.

The state of spectroscopy in 1890-1894 is beautifully summarized in the book
entitled \textit{Astronomical Spectroscopy} written in German in 1890 by Julius
Scheiner at Potsdam Observatory but translated into English in 1894 by Edwin
Frost \citep{1890spectral.book.S}.  This was the first textbook on the subject,
and as mentioned above, there is no discussion in this book on how to optimize
the performance of a spectrograph by selecting the spectrograph camera optics.
Spectroscopy was rapidly evolving in this period of time as evidenced by papers
published in the \emph{Astrophysical Journal} (e.g. \cite{1895ApJ.....2..370W}
and references therein).

Given the leadership shown by Scheiner in astronomical spectroscopy, perhaps it
is not surprising that he was the first to photographically detect the
absorption line spectrum of the Andromeda nebula \citep{1899ApJ.....9..149S}.
His detection of Andromeda, and an even earlier but less certain one by
\cite{1899.Huggins..An.Atl}, were both referenced by \cite{1909PASP...21..138F}
when Fath published the results of his Lick Observatory Ph.D. thesis research.
In this early work Fath detected photographically the spectra of six spiral
nebulae, with Andromeda being one of the six.

The theme of this brief historical contribution is to highlight the necessity of
using a fast spectrograph camera if the aim is to detect the absorption line
spectrum of spiral galaxies.  In his spectroscopic detection of Andromeda,
\cite{1899ApJ.....9..149S} used an \textit{f}/3 mirror system (most likely in
the configuration of a Zollner ocular spectrograph as described on page 81 of
his textbook).  \cite{1909PASP...21..138F} used an \textit{f}/3 spectrograph
camera lens in his work at Lick Observatory and later an \textit{f}/2 lens when
he worked at Mt. Wilson, and Vesto Melvin (``V.M.") Slipher employed an
\textit{f}/2.5 commercial camera lens in the spectrograph he used to detect
spiral galaxy spectra for measuring Doppler shifts.

\section{Brashear Spectrographs for Lick and Lowell Observatories}

In the early 1890s W.~W. Campbell, one of the great stellar spectroscopists of
the late 19\textsuperscript{th} and early 20\textsuperscript{th} centuries,
designed Lick Observatory's Mills spectrograph in conjunction with its
manufacturer, the John A. Brashear Company of Pittsburgh.  The Mills
spectrograph arrived at Lick Observatory in 1894 and was set to work collecting
photographic spectra of stars.  Soon after the Mills spectrograph was delivered,
Percival Lowell approached John Brashear to build a spectrograph for the Lowell
Observatory 24-inch refractor.  These two spectrographs share many of the same
design characteristics.  In particular, the spectrograph cameras were
\textit{f}/14 in both instruments, and both accommodated three prisms allowing
them to work at high spectral dispersion.

Percival Lowell hired V.M. Slipher in 1901, fresh out of Indiana University, to
commission the Brashear spectrograph on the Lowell Observatory 24-inch refractor
\citep{1980BioMemoirsNAS.H, 1994Smith...Slipher}.  Slipher's first scientific
assignment was to measure the rotation rate of the planet Venus, and before
starting this more difficult task, Slipher confirmed his techniques by obtaining
spectra with the Brashear spectrograph of the planets Jupiter and Saturn.  Since
all three planets--Jupiter, Saturn and Venus--are high surface brightness
objects, the Brashear spectrograph was able to make the necessary detections in
the original design configuration.   Slipher successfully determined an
upper-limit to Venus' rotation as no rotation could actually be detected.  He
finished this work by 1903.

Throughout the period from 1901 to 1910, V.M. Slipher interspersed his
investigations of planetary rotation with spectroscopic and radial velocity
measurements of stars.  One of his papers was a systematic investigation of
radial velocity standard stars with the Brashear spectrograph
\citep{1905ApJ....22..318S}.  During this same period he published approximately
30 papers in journals and in the Lowell Observatory Bulletin.

As documented in their correspondence, Percival Lowell began to encourage V.M.
Slipher starting in 1909 to work in an entirely new area of research:  to
determine the spectroscopic properties of the light coming from the outer parts
of spiral nebulae.  As reported by \cite{1994Smith...Slipher}, Lowell was
motivated by the idea that spiral nebulae might resemble proto-solar systems and
that the spectra from the outer regions of spirals might resemble the spectra of
the giant outer planets.  All observations of the outer regions of spirals in
this era were destined to be unsuccessful.  Venus and Jupiter have an
approximate surface brightness of 1.2 mag/arcsec\textsuperscript{2} and 5.4
mag/arcsec\textsuperscript{2}, respectively.  Bright galaxy nuclei have a
surface brightness in the range of 10 to 12 mag/arcsec\textsuperscript{2}, and
yet the outer parts of spirals (the region of Percival Lowell's interest) have a
surface brightness in the range of 21 mag/arcsec\textsuperscript{2}.  The 15
magnitude difference in surface brightness between Jupiter and the outer parts
of spirals is a factor of 1 million in terms of flux received at the
photographic plate.  The surface brightness difference between Jupiter and the
nuclei of spirals--which Scheiner, Fath and Slipher all detected--was a factor
of several hundred.

\section{Fath and Slipher Extend the Work of Scheiner}

Percival Lowell was not the only astronomer who had an interest in the spectra
of spiral nebulae.  Others speculated that spiral nebulae were galaxies of stars
and that their spectra could reveal key evidence in this regard.  By 1907 W.~W.
Campbell at Lick Observatory encouraged Ph.D. student Edward Fath to begin
spectroscopic observations of spiral nebulae and globular clusters with the
36-inch Crossley reflector.  Fath's spectrograph went together quickly because
it was assembled in a wooden box.  It was so crude that adding spectral
comparison lines to any spectrum meant that the spectrograph had to be
completely removed from the telescope.  These characteristics precluded Fath
from measuring reliable radial velocities.  But much to his credit, Fath's new
spectrograph at Lick Observatory used a fast spectrograph camera lens:  f/3.04.
By 1908 he had succeeded in detecting spectral lines (both absorption and
emission) in a few galaxies, and by 1909 he had published his first results
(those based on his Ph.D. thesis work).  Eventually, Fath published three papers
documenting observations for a total of ten galaxy spectra
\citep{1909PASP...21..138F, 1911ApJ....33...58F, 1913ApJ....37..198F}.

Those who read Fath's original work from 1909 can recognize that his
observations raised as many new questions as they solved.  He did see absorption
lines from the constituent stars in the spirals, but the spectra of spirals also
showed emission lines from Seyfert galaxy activity.   The original data are
described in \cite{1909PASP...21..138F}.  These stand on their own and are easy
to understand in a modern context, but his interpretation of his own
observations \citep{1909LicOB...5...71F} reflects the great uncertainty that
existed in that era regarding the nature of the spiral nebulae.

After graduating with his Ph.D. degree, Fath accepted a position at Mt. Wilson
Observatory where he continued his work on spiral nebulae by building another
spectrograph for the newly completed Mt. Wilson 60-inch telescope.  His second
spectrograph was no more robust than his first.  Slipher and Fath met for the
first time at Mt. Wilson Observatory in August, 1910, approximately one year
after Fath's first paper on spiral nebulae had been published.  They were both
attending a large astronomical gathering entitled the
\textit{4\textsuperscript{th} Conference of the International Union for
Cooperation in Solar Research}.  Soon after they met at this conference, they
began an intermittent correspondence that lasted three years.  Based on the
correspondence between Fath and Slipher (available in the Lowell Observatory
Archives) and on the timing of Slipher's subsequent work, it appears that their
meeting at Mt. Wilson helped to spur Slipher into action.  With the mechanical
assistance of Stanley Sykes, Slipher modified the Brashear spectrograph allowing
it to accept a fast spectrograph camera lens.  Because Fath's spectrographs were
neither rigid nor stable enough to provide Doppler shifts, a perfect opportunity
was open for V.M. Slipher to apply to the spiral nebula problem everything he
had learned about Doppler shifts and the Brashear spectrograph during the
previous ten years.   Slipher purchased an \textit{f}/2.5 commercial camera lens
from Voigtlander and installed it in the Brashear spectrograph.  Figure 1 shows
the new equipment in its final assembled form.

\begin{figure}[!ht]
\begin{center}\plotone{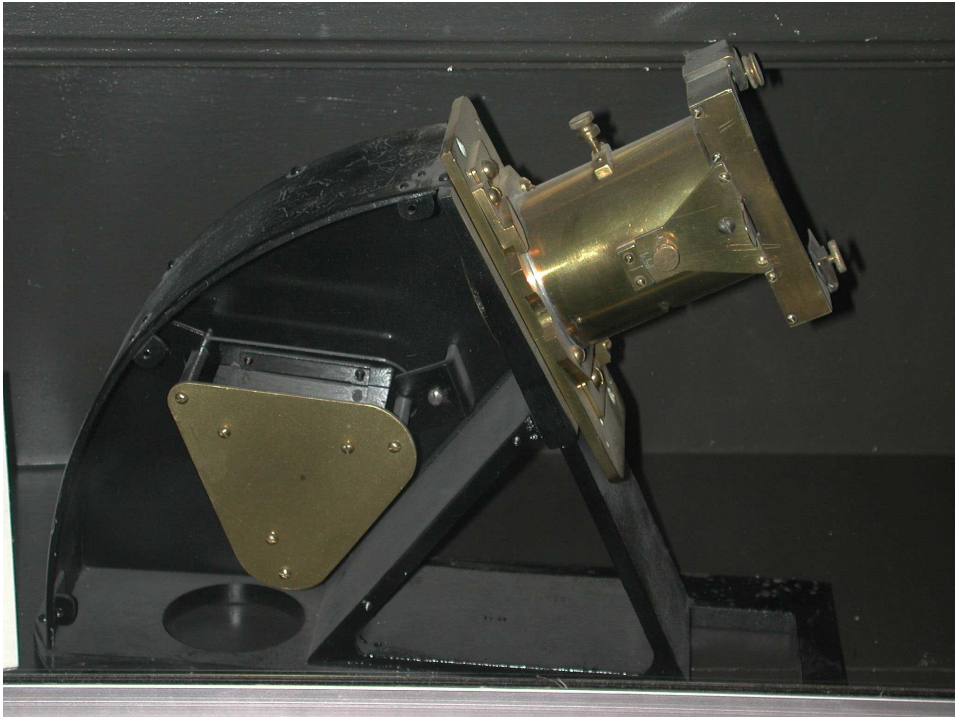}
\caption{The collimated beam from the Brashear spectrograph enters from below
through the hole on the lower left, passes through the prism (visible under the
large triangular plate), and proceeds to the \textit{f}/2.5 Voigtlander lens
that is mounted inside the large brass cylinder in the upper right.  The
photographic plate holder is held under two mounting screws on the upper right
hand side at the focal plane of the Voigtlander lens.} \end{center}
\end{figure}

Slipher's first target was the Andromeda nebula, and his first opportunity to
use the new spectrograph camera came in December 1910.  His early spectra showed
some absorption lines, but the results were not optimal.  In a letter to Fath
dated February 8, 1911, Slipher \footnote{\cite{Slipher1911}}
estimates that the sky was so poor during these
exposures that his equivalent total exposure time in 1910 amounted to no more
than four hours.  V.M. Slipher tried again in the fall of 1911 and found that
his second effort was also unsuccessful because he had used a prism with too low
a dispersion and a slit that was too narrow (relative to the grain in the
photographic emulsion).  By fall 1912, he was using a higher dispersion prism
(i.e. a prism of denser glass) and a wider slit.  Both the 1911 and the 1912
configurations gave the same spectral resolution, but the latter combination
produced a better image of the slit on the photographic emulsion
\citep{1913LowOB...2...56S}.  It was at that point (September 17, 1912) that on
an exposure lasting one full night he obtained his first spectrum of Andromeda
with an image that could be measured for the Doppler shift.  He followed up
rather quickly with more exposures, the first pair each lasting two nights
(November 15+16 and then December 3+4) and the final one lasting three nights
(December 29+30+31).  All of the spectra taken in late 1912 yielded measurable
Doppler shifts.

In his early papers on Andromeda, Slipher made no mention of the slit width he
used in the spectrograph.  But in \cite{1917PAPhS..56..403S} he states that his
prism worked at 140 {\AA} mm$^{-1}$ and that the equivalent slit width was 0.06 mm.  I
interpret this to mean that the slit projects onto the plate with a width of
0.06 mm, and working backwards with the optical properties of the spectrograph
and the 24-inch Lowell refractor, this means that on the sky the slit width was
8.25 arcsec.  \cite{1909PASP...21..138F} states that on the Crossley reflector,
he used a slit width of 0.15 mm, and this converts to an angular width of 5.8
arcsec.  So both the Fath and Slipher instruments were working as nebular
spectrographs at very low spectral dispersion with wide slits.  Only in this way
could they attain adequate signal-to-noise ratios on the final photographic
image.

\section{The Defining Properties of Nebular Spectrographs}

Although Scheiner and Fath were the first to use short focal length camera
lenses in their spectrographs thereby making the absorption lines of spiral
nebulae detectable, V.M. Slipher went a step further in an effort to determine
what parameters--for both the telescope and the spectrograph--were responsible
for optimizing the detection of low surface brightness objects.  Slipher's
conclusions first appeared in a letter to Fath dated February 8, 1911:

\begin{quote}
the ratio of the aperture to focus of the telescope objective has nothing to
do with its usefulness for spectrum work on the extended surface \ldots
Intensity of the image on the slit of course does not count but (the)
intensity of spectrum on the sensitive surface does and it is only the camera
that determines this.
\end{quote}

In this correspondence Slipher asks Fath directly whether spectra of spiral
nebulae taken with the Mt. Wilson 60-inch telescope are significantly better than
those taken with the 36-inch Crossley reflector, but Fath was unable to give any
clear evidence one way or the other.  He simply replied that the two
spectrographs he used were sufficiently different that he could not tell.  Of
course, Slipher was working with the Lowell Observatory 24-inch refractor with a
long focal length objective, and this matter was of key importance to him.

V.M. Slipher's design specifications for nebular spectrographs were repeated in
a very clear manner in all three of his papers on the Doppler shifts in galaxy
spectra:  \cite{1913LowOB...2...56S}, \cite{1915..Spectroscop.S},
\cite{1917PAPhS..56..403S}.  These ideas became of central importance to the
later work of Milton \cite{1931ApJ....74...35H}, work that provided the
foundation for Edwin Hubble's extension of the velocity-distance relation beyond
the initial forty-one Doppler shifts determined by V.M. Slipher (those 41 were
originally published as a complete list in \cite{1923mtr..book.....E}).  It is
emphatically true that without Slipher's initial 41 Doppler shifts and without
his clearly stated design principles for nebular spectrographs, Humason and
Hubble's work on the velocity-distance relation would have suffered a
considerable delay.

By the 1930s, the key role of the spectrograph camera's \textit{f}/ratio was
recognized by the staff of the Mt. Wilson Observatory.  The first spectrograph
given to Humason for redshift work used an \textit{f}/1.43 camera lens.  Later
Humason had access to a series of extremely fast camera lenses with
\textit{f}/0.59 built specifically for the Humason-Hubble redshift program
\citep{1930ApJ....72...59R}.  The ease with which Humason pushed to higher
redshifts on the Mt. Wilson 100-inch telescope, far beyond those of Slipher, was
almost entirely dependent on the speed and excellent optical quality of the
Rayton spectrograph camera lens and had little to do with the 100-inch telescope
aperture.

A modern re-statement of Slipher's nebular spectrograph design concepts was made
by Mt. Wilson Observatory's Ira Bowen in 1952 when he wrote in a more complete
way (for both nebular and stellar sources) how the ``speed" of a spectrograph
depends on key parameters like the diameter of the telescope objective, the
\textit{f}/ratio of the spectrograph camera lens, and (for stars) the slit width
relative to the seeing disk \citep{1952ApJ...116....1B}.  For a true nebular
spectrograph, exactly as V.M. Slipher originally stated, Bowen found the
``speed" of detection (i.e. the energy cm$^{-2}$ s$^{-1}$ received at the
detector) is completely independent of the diameter and focal length of the
telescope objective and can be written as follows: \begin{equation}
``speed" = \textup{(object surface brightness) } w' P /  F\textsuperscript{2}
\end{equation}
where $w'$ is the projected slit width in millimeters at the detector, $P$
is the spectral dispersion in {\AA}ngstroms mm$^{-1}$ at the detector, and $F$ is the focal
ratio of
the spectrograph camera lens.  For example, if we assume that $w' P$ (the number
of {\AA}ngstroms sampled by the spectrograph slit at the photographic plate) was the
same in Slipher's nebular spectrograph as it was in Humason's spectrograph with
the Rayton lens, the relative ``speed" of detection between their two systems
when working on the same object is the square of the ratio of the two cameras
\textit{f}/ratios: \begin{equation}
(2.5 / 0.59)\textsuperscript{2}  =  18.
\end{equation}
So it is not surprising that the Humason-Hubble team pushed deeper into the
Universe when they confirmed in the 1930s the linear velocity distance
relation.

Those who want to see a more complete discussion of the speed of spectrographs
and their key design parameters might look at the somewhat updated description
in \cite{1964aste.book...34B} or at monographs on instrumentation design like
\textit{Astronomical Optics} by \cite{1987sdap.book.....S}.  The use of the
symbols $w'$ and $P$ above are taken from \cite{1987sdap.book.....S}, and I have
omitted from the simple equations above several parameters, for example the
end-to-end optical system efficiency and the ``anamorphic magnification" in the
spectrograph beam, because these two parameters will be somewhat similar from
one optical system to the other.

\acknowledgements I thank Ms. Lauren Amundson for providing
scanned copies of the
correspondence between Edward Fath and V.M. Slipher from the Lowell Observatory
Archives.  I thank the two referees, Joseph Tenn and John Hearnshaw, for their
help in improving the clarity and contents of this contribution.  Generous
thanks are also due to both the Scientific Organizing Committee and the Local
Organizing Committee for making this conference a success and for following
through with the publication of the proceedings.

\bibliography{thompson}

\begin{thebibliography}{}
\expandafter\ifx\csname natexlab\endcsname\relax\def\natexlab#1{#1}\fi
\expandafter\ifx\csname url\endcsname\relax
  \def\url#1{\texttt{#1}}\fi
\expandafter\ifx\csname urlprefix\endcsname\relax\def\urlprefix{URL }\fi
\providecommand{\eprint}[2][]{\url{#2}}

\bibitem[{{Bowen}(1952)}]{1952ApJ...116....1B}
{Bowen}, I.~S. 1952, {The Spectrographic Equipment of the 200-inch Hale
  Telescope}, \apj, 116, 1

\bibitem[{{Bowen}(1964)}]{1964aste.book...34B}
--- 1964, {Spectrographs, in Astronomical Techniques, ed. W.~A. Hiltner}
  (University of Chicago Press: Chicago), chap.~2, 34

\bibitem[{{Eddington}(1923)}]{1923mtr..book.....E}
{Eddington}, A.~S. 1923, {The Mathematical Theory of Relativity} (Cambridge
  University: Cambridge, England)

\bibitem[{{Fath}(1909{\natexlab{a}})}]{1909PASP...21..138F}
{Fath}, E.~A. 1909{\natexlab{a}}, {The Spectra of Some Spiral Nebul{\ae} and
  Globular Star Clusters}, \pasp, 21, 138

\bibitem[{{Fath}(1909{\natexlab{b}})}]{1909LicOB...5...71F}
--- 1909{\natexlab{b}}, {The Spectra of Some Spiral Nebul{\ae} and Globular
  Star Clusters}, Lick Observatory Bulletin, 5, 71

\bibitem[{{Fath}(1911)}]{1911ApJ....33...58F}
--- 1911, {The Spectra of Some Spiral Nebul{\ae} and Globular Star Clusters},
  \apj, 33, 58

\bibitem[{{Fath}(1913)}]{1913ApJ....37..198F}
--- 1913, {The Spectra of Spiral Nebulae and Globular Star Clusters. Third
  Paper}, \apj, 37, 198

\bibitem[{{Hoyt}(1980)}]{1980BioMemoirsNAS.H}
{Hoyt}, W.~G. 1980, {Vesto Melvin Slipher 1875-1969}, vol.~52 (Biographical
  Memoirs, National Academy of Sciences)

\bibitem[{{Huggins}(1899)}]{1899.Huggins..An.Atl}
{Huggins}, W. 1899, An Atlas of Representative Stellar Spectra from lambda 4870
  to lambda 3300 (Publications of Sir William Huggins Observatory: London,
  England)

\bibitem[{{Humason}(1931)}]{1931ApJ....74...35H}
{Humason}, M.~L. 1931, {Apparent Velocity-Shifts in the Spectra of Faint
  Nebulae}, \apj, 74, 35

\bibitem[{{Rayton}(1930)}]{1930ApJ....72...59R}
{Rayton}, W.~B. 1930, {Two High-Speed Camera Objectives for Astronomical
  Spectrographs}, \apj, 72, 59

\bibitem[{{Scheiner}(1890)}]{1890spectral.book.S}
{Scheiner}, J. 1890, Die Spectralanalyse der Gestirne. trans: E.~B. Frost,
  Astronomical Spectroscopy (Ginn and Company: Boston)

\bibitem[{{Scheiner}(1899)}]{1899ApJ.....9..149S}
--- 1899, {On the Spectrum of the Great Nebula in Andromeda}, \apj, 9, 149

\bibitem[{{Schroeder}(1987)}]{1987sdap.book.....S}
{Schroeder}, D.~J. 1987, {Astronomical Optics} (Academic Press, Inc.: San
  Diego, CA)

\bibitem[{{Slipher}(1905)}]{1905ApJ....22..318S}
{Slipher}, V.~M. 1905, {Observations of Standard Velocity Stars with the Lowell
  Spectrograph}, \apj, 22, 318

\bibitem[{{Slipher}(1911)}]{Slipher1911}
--- 1911, {Letter to Fath, 8 February 1911}. Letter

\bibitem[{{Slipher}(1913)}]{1913LowOB...2...56S}
--- 1913, {The Radial Velocity of the Andromeda Nebula}, Lowell Observatory
  Bulletin, 2, 56

\bibitem[{{Slipher}(1915)}]{1915..Spectroscop.S}
--- 1915, {Spectroscopic Observations of Nebulae}, {Popular Astronomy}, 23, 21

\bibitem[{{Slipher}(1917)}]{1917PAPhS..56..403S}
--- 1917, {Nebulae}, Proceedings of the American Philosophical Society, 56, 403

\bibitem[{{Smith}(1994)}]{1994Smith...Slipher}
{Smith}, R.~W. 1994, {Redshift and Gold Medals in W.~L. Putnam's The Explorers
  of Mars Hill : A Centennial History of Lowell Observatory, 1894-1994}
  (Phoenix Publishing: West Kennebunk, Maine), chap.~4, 41

\bibitem[{{Wadsworth}(1895)}]{1895ApJ.....2..370W}
{Wadsworth}, F.~L.~O. 1895, {The Modern Spectroscope. XIV. Fixed-Arm Concave
  Grating Spectroscopes}, \apj, 2, 370

\end{thebibliography}

\end{document}